\documentclass{article}
\usepackage{times}
\usepackage{amsfonts}
\usepackage{graphicx}
\begin{document}
\noindent
{\Large   ON THE COSMOLOGICAL CONSTANT OF FLAT FLRW SPACETIME}
\noindent

\vskip.5cm
\noindent
{\bf P. Fern\'andez de C\'ordoba}$^{a,1}$,  {\bf R. Gallego Torrom\'e}$^{b,2}$,
{\bf S. Gavasso}$^{a,3}$ and {\bf  J.M. Isidro}$^{a,4}$\\
$^{a}$Instituto Universitario de Matem\'atica Pura y Aplicada,\\ Universitat Polit\`ecnica de Val\`encia, Valencia 46022, Spain\\
$^{b}$Department of Mathematics, Faculty of Mathematics, Natural Sciences\\ and Information Technologies, University of Primorska, Koper, 
Slovenia\\
$^{1}${\tt pfernandez@mat.upv.es}, $^{2}${\tt rigato39@gmail.com},\\ 
$^{3}${\tt sgavas@alumni.upv.es}, $^{4}${\tt joissan@mat.upv.es}\\

\noindent
\vskip.5cm
\noindent
{\bf Abstract}  We consider an exponentially expanding, flat, Friedmann-Lema\^{\i}tre-Robertson-Walker (FLRW) Universe filled with a free Schroedinger field. The probability fluid of the latter is used to mimic the cosmological fluid (baryonic plus dark  matter), thus providing the matter density and pressure terms in the corresponding Friedmann-Lema\^{\i}tre equations. We first obtain the eigenfunctions of the Laplacian operator on flat FLRW space. A quantum operator qualifying as a cosmological constant is defined to act on the Schroedinger field. We then compute the matrix representing the cosmological constant in the basis of Laplacian eigenfunctions. For an estimate of the orders of magnitude involved it suffices to determine the expectation values of this operator. The expectation value that best fits the experimentally measured value of the cosmological constant allows us to identify the quantum state of the Schroedinger field that best represents the matter contents (baryonic and dark) of the current Universe. Finally, the operator inverse (modulo dimensional factors) to the one representing the cosmological constant provides a measure of the gravitational Boltzmann entropy of the Universe. We compute its matrix in the basis of Laplacian eigenfunctions and verify that the expectation values of this entropy operator comply with the upper bound set by the holographic principle.


\section{Introduction}\label{einfuehrung}

\subsection{The cosmological fluid as a Newtonian fluid}\label{cfasanwt}

Our present understanding of the Universe describes it fairly accurately by means of a flat, Friedmann-Lema\^{\i}tre-Robertson-Walker (FLRW) geometry \cite{WEINBERG2}, with a tiny but positive cosmological constant $\Lambda$ driving its accelerated expansion \cite{PERLMUTTER, RIESS}. At the same time, while spacetime requires general relativity for its precise comprehension, in many situations the matter content within it (baryonic plus dark) appears to be reasonably well described by Newtonian dynamics. 

The fact that the cosmological fluid often behaves nonrelativistically allows to regard it as a Newtonian fluid, generally an ideal fluid. That means it is subject to the continuity equation and the Euler equation:
\begin{equation}
\frac{\partial\rho}{\partial t}+\nabla\cdot\left(\rho{\bf v}\right)=0,\qquad \frac{\partial{\bf v}}{\partial t}+\left({\bf v}\cdot\nabla\right){\bf v}+\frac{1}{\rho}\nabla p-{\bf F}=0.
\label{oule}
\end{equation}
Above, $\rho$ is the volume density of fluid matter, $p$ its pressure,  ${\bf v}$ the velocity field and ${\bf F}$ the external force acting per unit mass of the fluid. In the presence of viscosity, the Euler equation is replaced by the Navier--Stokes equation, but we will disregard this possibility. A comprehensive discussion of fluids within FLRW spacetimes is given in the review \cite{BAMBA}.

Besides the cosmological fluid, there is another nonrelativistic framework that can be bijectively mapped into an ideal fluid dynamics, namely the Schroedinger quantum mechanics of a single scalar particle of mass $M$. Following Madelung \cite{MADELUNG} one factorises the wavefunction $\Psi$ into amplitude and phase as
\begin{equation}
\Psi=\exp\left(S+\frac{{\rm i}}{\hbar}I\right).
\label{fatto}
\end{equation}
The phase $\exp({\rm i}I/\hbar)$ is the complex exponential of the classical--mechanical action integral $I$, and we have written the amplitude as the exponential of the real, dimensionless quantity $S$. Away from the zeroes of $\Psi$, the above factorisation is always possible. The ideal fluid at work here is the {\it quantum probability fluid}\/. It has a velocity field ${\bf v}$ and a volume density of matter $\rho$ defined by
\begin{equation}
{\bf v}=\frac{1}{M}\nabla I, \qquad \rho={\rm e}^{2S}=\vert\Psi\vert^2.
\label{ohsolemio}
\end{equation}
Substituting the factorisation (\ref{fatto}) into the time--dependent Schroedinger equation with a potential $V$ one arrives at the set of two equations
\begin{equation}
\frac{\partial S}{\partial t}+\frac{1}{M}\nabla S\cdot\nabla I+\frac{1}{2M}\nabla^2I=0
\label{konntt}
\end{equation}
and
\begin{equation}
\frac{\partial I}{\partial t}+\frac{1}{2M}(\nabla I)^2+V+Q=0.
\label{quuve}
\end{equation}
Eq. (\ref{konntt}) is an equivalent rewriting of the usual continuity equation in (\ref{oule}), while (\ref{quuve}) is the {\it quantum Hamilton--Jacobi equation}\/. It contains the external potential $V$ present in the Schroedinger equation, plus an additional term $Q$ known as the {\it quantum potential}\/:
\begin{equation}
Q=-\frac{\hbar^2}{2M}\left[\left(\nabla S\right)^2+\nabla^2 S\right].
\label{nanunana}
\end{equation}
The quantum Hamilton--Jacobi equation reduces to its classical counterpart in the limit $\hbar\to 0$, when the quantum potential $Q$ vanishes. 

Since we are dealing with a Newtonian fluid, we need an Euler equation. Following Madelung \cite{MADELUNG} it is provided by the gradient of Eq. (\ref{quuve}):
\begin{equation}
\frac{\partial{\bf v}}{\partial t}+\left({\bf v}\cdot\nabla\right){\bf v}+\frac{1}{M}\nabla Q+\frac{1}{M}\nabla V=0.
\label{kblm}
\end{equation}
Comparison between Eqs. (\ref{kblm}) and (\ref{oule}) produces a one--to--one correspondence between the quantum probability fluid and an Euler fluid, according to the table below:
\begin{equation}
\begin{tabular}{| c | c | c |}\hline
& Euler &  Madelung  \\ \hline
volume density& $\rho$ & $\exp(2S)$   \\ \hline
velocity & ${\bf v}$ & $\nabla I/M$   \\ \hline
pressure term & $\nabla p/\rho$ & $\nabla Q/M$   \\ \hline
external forces & ${\bf F}$ & $-\nabla V/M$   \\ \hline
\end{tabular}
\label{tabla}
\end{equation}
We conclude that, given a Newtonian cosmological fluid, one can use nonrelativistic quantum mechanics to describe it. At the very least, this analysis in terms of the probability fluid corresponding to the wavefunction $\Psi$ is equivalent to the original one. We will see, moreover, that this alternative description offers a number of advantages. 

{}From now on we set $c=1=\hbar$. The Boltzmann constant $k_B$ will also have a role to play; we will likewise set $k_B=1$. Specifically, the quantity $2S$ in Eq. (\ref{tabla}) will turn out to be the dimensionless entropy ${\cal S}$ defined in Eq. (\ref{trumptheloser}), {\it i.e.}\/, ${\cal S}/(2k_B)=S$.\footnote{Note the different typographies for ${\cal S}$ and $S$.}

\subsection{A Newtonian cosmological fluid within a relativistic spacetime}\label{ncfasarltvspt}

We have in previous publications made use of the bijection (\ref{tabla}) to analyse the Newtonian cosmological fluid within the Newtonian spaces 
$\mathbb{R}\times\mathbb{R}^3$ (flat space) \cite{PLANO}, $\mathbb{R}\times\mathbb{H}^3$ (hyperbolic space) \cite{HIPERBOLICO} and $\mathbb{R}\times\mathbb{S}^3$ (spherical space) \cite{ESFERICO}, the real line $\mathbb{R}$ standing for the time axis in all three cases. In all of them gravity has been assumed so weak that it could be dealt with nonrelativistically, {\it i.e.}\/, the corresponding spaces were fixed backgrounds with no dynamics at all. Useful though these examples have been as warmup exercises, they fall short of treating spacetime dynamically.

In the present paper we address the need for treating spacetime dynamically according to general relativity. At the same time we continue to regard the cosmological fluid as a Newtonian fluid described by a nonrelativistic, quantum--mechanical wavefunction $\Psi$. Thus the matter contents of the Universe will be nonrelativistic, but the spacetime will be relativistic (flat FLRW space). In particular, the latter will be governed by the Friedmann--Lema\^{\i}tre equations. That this apparent mismatch of simultaneously having relativistic and nonrelativistic approaches within one given model causes no incompatibility will be demonstrated next. 

Our starting point is the flat FLRW metric expressed in comoving coordinates \cite{WEINBERG2},
\begin{equation}
{\rm d}s^2={\rm d}t^2-a^2(t)\left({\rm d}r^2+r^2{\rm d}\Omega^2\right), \qquad {\rm d}\Omega^2={\rm d}\theta^2+\sin^2\theta\,{\rm d}\phi^2.
\label{metrik}
\end{equation}
In the presence of an ideal fluid of density $\rho$ and pressure $p$, the corresponding Friedmann--Lema\^{\i}tre equations read
\begin{equation}
3\,\frac{\dot a^2}{a^2}=\kappa\rho+\Lambda,\qquad 2\,\frac{\ddot a}{a}+\frac{\dot a^2}{a^2}=-\kappa p+\Lambda,
\label{kklar}
\end{equation}
where $\kappa=8\pi G$ and a dot denotes ${\rm d}/{\rm d}t$. Given the extreme smallness of $\Lambda$ we may take 
\begin{equation}
3\,\frac{\dot a^2}{a^2}=\kappa\rho,\qquad 2\,\frac{\ddot a}{a}+\frac{\dot a^2}{a^2}=-\kappa p
\label{kklarx}
\end{equation}
as a valid starting point; $\Lambda$ will be computed presently as a correction to Eq. (\ref{kklarx}).

Next we argue why the relativistic fluid on the right--hand side of the Friedmann equations (\ref{kklarx}) can become Newtonian without spoiling the consistency of the formalism. It is perfectly consistent to treat spacetime relativistically but the matter within it nonrelativistically because the structure of the FLRW metric (\ref{metrik}) allows it; let us explain how this comes about.

The FLRW Laplacian $\square_{\rm FLRW}$ is given in Eq. (\ref{laplos}) below. Now the scale factor $a(t)$ does not depend on the space variables; furthermore it multiplies a space metric that coincides with its nonrelativistic counterpart. It is usual to set $\dot a(t)/a(t)=H(t)$; as a first approximation we can assume $H(t)$ to be time--independent (Hubble's constant $H_0$). Then
\begin{equation}
a(t)=\exp(H_0t),
\label{espansione}
\end{equation}
in good agreement with experimental data \cite{PERLMUTTER, RIESS} suggesting that the Universe currently finds itself in a period of exponential expansion  due to the increasing dominance of the vacuum energy (for reviews see, {\it e.g.}\/, refs. \cite{CARROLL, PEEBLES, WEINBERG1}). Under the assumption (\ref{espansione}), the operator $\square_{\rm FLRW}$ of Eq. (\ref{laplos}) simplifies to the operator $\square_{H_0}$ of Eq. (\ref{laplass}). When solving the eigenvalue equation for $\square_{H_0}$,  the 3--dimensional Laplacian $\nabla^2$ present in (\ref{laplass}) can be used to define a nonrelativistic, free Hamiltonian operator for the matter contents of the Universe. This is in accordance with the fact that the cosmological fluid can often be regarded as nonrelativistic. 

We observe that the ansatz (\ref{espansione}), when substituted into Eq. (\ref{kklarx}), leads to 
\begin{equation}
3H_0^2=\kappa\rho, \qquad 3H_0^2=-\kappa p.
\label{essai}
\end{equation}
Two consequences are the constancy of $\rho$ across spacetime, and the equality $p=-\rho$. A constant density $\rho$ is in agreement with the cosmological principle, but a negative pressure fails to qualify as a valid equation of state for baryonic matter; instead $p=-\rho$ is the equation of state for the cosmological constant \cite{WEINBERG2}. We will later elaborate on the implications that a constant density $\rho$ has on the Schroedinger wavefunction $\Psi$ describing the cosmological fluid.

\subsection{Some relevant numbers}\label{numeri}

Let $M_{\rm tot}$  denote the total mass/energy contents of the Universe, $H_0$ the Hubble constant, $R_U$ the radius of the observable Universe, and $\Lambda$ the cosmological constant. We collect their numerical values from ref. \cite{PLANCK}:
\begin{equation}
M_{\rm tot}=3.0\times 10^{54}\,{\rm kg}, \qquad H_0=2.2\times 10^{-18}\, {\rm s}^{-1}, \qquad R_U=4.4\times 10^{26}\, {\rm m}.
\label{mase}
\end{equation}
{}For the cosmological constant we have
\begin{equation}
\Lambda=1.1\times 10^{-52}\, {\rm m}^{-2}, \qquad {\rm or} \qquad \Lambda=2.9\times 10^{-122}\; {\rm natural}\; {\rm units}.
\label{dosele}
\end{equation}
Specifically, $M_{\rm tot}$ is the sum of baryonic $M_{\rm bar}$, dark matter $M_{\rm dm}$ and dark energy $M_{\rm de}$ contributions: $M_{\rm tot}=M_{\rm bar}+M_{\rm dm}+M_{\rm de}$. We have the percentage fractions 
\begin{equation}
M_{\rm bar}=5M_{\rm tot}/100,\qquad M_{\rm dm}=27M_{\rm tot}/100,\qquad M_{\rm de}=68M_{\rm tot}/100.
\label{unitats}
\end{equation}
The previous numbers are accurate enough for our purposes; we will neglect indeterminacies such as the Hubble tension \cite{ELIZALDE1} and statistical errors. Contributions from radiation are negligible compared to those from matter. 

We have proposed a mechanism whereby the cosmological fluid is described by the probability fluid of a particle satisfying the nonrelativistic Schroedinger equation. For the (as yet undetermined) Schroedinger mass $M$ we will take
\begin{equation}
M=M_{\rm bar}+M_{\rm dm},
\label{massa}
\end{equation}
while dark energy will be accounted for by a quantum operator $\Lambda$ acting on $\Psi$, to be constructed in what follows.\footnote{We use the same notation $\Lambda$ to refer to the numerical value of the cosmological constant, and to the quantum operator representing it in our model.}

\section{The FLRW Laplacian}\label{flrwlaplass}

\subsection{Eigenfunctions and spectrum}\label{egfcspt}

The Laplacian operator $\square_{\rm FLRW}$ with respect to the FLRW metric (\ref{metrik}) reads
\begin{equation}
\square_{\rm FLRW}=3\,\frac{\dot a(t)}{a(t)}\frac{\partial}{\partial t}+\frac{\partial^2}{\partial t^2}-\frac{1}{a^2(t)}\nabla^2,
\label{laplos}
\end{equation}
where $\nabla^2$ denotes the 3--dimensional spatial Laplacian
\begin{equation}
\nabla^2=\frac{1}{r^2}\frac{\partial}{\partial r}\left(r^2\frac{\partial}{\partial r}\right)+\frac{1}{r^2}\left[\frac{1}{\sin\theta}\frac{\partial}{\partial\theta}\left(\sin\theta\frac{\partial}{\partial\theta}\right)+\frac{1}{\sin^2\theta}\frac{\partial^2}{\partial\phi^2}\right].
\label{esfelaplas}
\end{equation}
Using the scale factor  (\ref{espansione}), the Laplacian (\ref{laplos}) becomes
\begin{equation}
\square_{H_0}=3H_0\frac{\partial}{\partial t}+\frac{\partial^2}{\partial t^2}-\exp\left(-2H_0t\right)\nabla^2.
\label{laplass}
\end{equation} 
In solving the Laplacian eigenvalue problem
\begin{equation}
\square_{H_0}\Psi_{\mu}(t,r,\theta,\phi)=\mu\Psi_{\mu}(t,r,\theta,\phi)
\label{eijen}
\end{equation}
by separation of variables, $\Psi(t,r,\theta,\phi)=T(t)\psi(r,\theta,\phi)$, we will require that the spatial piece $\psi(r,\theta,\phi)$ belong to the Hilbert space of square--integrable functions on the spatial hypersurface $t=t_0$ within FLRW spacetime. As usual the scalar product is given by
\begin{equation}
\langle\psi_1\vert\psi_2\rangle=\int_0^{\infty}{\rm d}r\,\int_0^{\pi}{\rm d}\theta\,\int_0^{2\pi}{\rm d}\phi\,r^2\sin\theta\,\psi_1^*\psi_2.
\label{skal}
\end{equation}
In so doing one arrives at the set of two equations
\begin{equation}
\nabla^2\psi=C\,\psi
\label{diez}
\end{equation}
and
\begin{equation}
\frac{{\rm d}^2T}{{\rm d}t^2}+3H_0\frac{{\rm d}T}{{\rm d}t}-\left(C\,{\rm e}^{-2H_0t}+\mu\right)\,T(t)=0,
\label{once}
\end{equation}
where $C$ is a separation constant. Imposing the boundary condition that the spatial wavefunction $\psi$ be nonsingular at $r=0$, a complete, orthonormal set of eigenfunctions of the spatial Laplacian $\nabla^2$ can be obtained as usual (see, {\it e.g.}\/, ref. \cite{LANDAU2}). The spectrum of allowed values for the separation constant $C$ is
\begin{equation}
C=-k^2, \qquad k\in(0,\infty),
\label{spettro}
\end{equation}
and the spatial eigenfunction $\psi_{klm}(r,\theta,\phi)$ is given by the product of a radial function $R_{kl}(r)$ times a spherical harmonic 
$Y_{lm}(\theta,\phi)$,
\begin{equation}
\psi_{klm}(r,\theta, \phi)=R_{kl}(r)\,Y_{lm}(\theta,\phi).
\label{doce}
\end{equation}
Now substitution of the continuous spectrum (\ref{spettro}) into Eq. (\ref{once}) yields
\begin{equation}
\frac{{\rm d}^2T}{{\rm d}t^2}+3H_0\frac{{\rm d}T}{{\rm d}t}+\left(k^2\,{\rm e}^{-2H_0t}-\mu\right)\,T(t)=0.
\label{trece}
\end{equation}
Two linearly independent solutions of the above are
\begin{equation}
T_{\mu k}^{(\pm)}(t)=\exp\left(-\frac{3}{2}H_0t\right)J_{\pm\frac{1}{2H_0}\sqrt{9H_0^2+4\mu}}\left(\frac{k}{H_0}{\rm e}^{-H_0t}\right),
\label{jotauno}
\end{equation}
where $J_p(z)$ is a Bessel function of the first kind. For regularity as $t\to\infty$ we pick the positive sign in (\ref{jotauno}) and denote $T^{(+)}_{\mu k}(t)$ more simply as $T_{\mu k}(t)$.

Altogether, the eigenvalue problem for the FLRW Laplacian $\square_{H_0}$ is solved by
\begin{equation}
\Psi_{\mu klm}(t,r,\theta,\phi)=T_{\mu k}(t)\psi_{klm}(r,\theta,\phi),\qquad \mu\in\mathbb{R}.
\label{fattoz}
\end{equation}
The $\psi_{klm}(r,\theta,\phi)$ form a complete, orthonormal set within the Hilbert space of stationary quantum states of a free particle within $\mathbb{R}^3$ as $k\in(0,\infty)$, $l\in\mathbb{N}$ and $m=-l,\ldots, l$.

\subsection{Relationship with Klein--Gordon and Schroedinger}\label{rltkgschd}

A particular instance of the eigenvalue equation (\ref{eijen}) is the case when 
\begin{equation}
\mu=-M^2,
\label{lotengo}
\end{equation}
with $M$ given by Eq. (\ref{massa}). Then (\ref{eijen}) becomes the Klein--Gordon equation on FLRW space,
\begin{equation}
\left(\square_{H_0}+M^2\right)\Psi=0,
\label{kage}
\end{equation}
and the index $p$ of the Bessel functions in (\ref{jotauno}) becomes, in view of $M\gg H_0$, 
\begin{equation}
p=\frac{1}{2H_0}\sqrt{9H_0^2-4M^2}\simeq\frac{{\rm i}M}{H_0}.
\label{indexvesel}
\end{equation}
Thus a Klein--Gordon field within flat FLRW space is described by the (closure of the linear span of the) wavefunctions
\begin{equation}
\Psi_{Mklm}(t,r,\theta,\phi)=\exp\left(-\frac{3}{2}H_0t\right)J_{{\rm i} \frac{M}{H_0}}\left(\frac{k}{H_0}{\rm e}^{-H_0t}\right)\psi_{klm}(r,\theta,\phi).
\label{fattox}
\end{equation}
The functions (\ref{fattox}) are simultaneous eigenfunctions of the Laplacian operators $\square_{H_0}$ and $\nabla^2$. Now
$-\nabla^2/(2M)$ is the nonrelativistic Hamiltonian ${\cal H}$ for a free particle, so the $\Psi_{Mklm}(t,r,\theta,\phi)$ also diagonalise ${\cal H}$, with the time--dependent piece playing a spectator role. In particular, the Hilbert space of stationary quantum states for a free, nonrelativistic, Schroedinger  fluid within an exponentially expanding FLRW space is obtained as the (closure of) the linear span of the $\psi_{klm}(r,\theta,\phi)$  in Eq. (\ref{fattox}).

\subsection{The subspace of radially symmetric quantum states}

The cosmological principle demands that we restrict our attention to spherically symmetric states, {\it i.e.}\/, to states with $l=0$, $m=0$.
Then the subspace of spherically symmetric quantum states is spanned by the subset of the wavefunctions (\ref{doce}) given by 
\begin{equation}
\psi_{k00}(r,\theta,\phi)=(4\pi)^{-1/2}R_{k0}(r), \qquad R_{k0}(r)=\sqrt{\frac{2}{\pi}}\,\frac{\sin(kr)}{r}, \quad k\in(0,\infty).
\label{espan}
\end{equation}
We will denote $\psi_{k00}$ more simply by $\psi_k$. These states are orthonormal with respect to the scalar product (\ref{skal}),
\begin{equation}
\langle\psi_k\vert\psi_{k'}\rangle=\delta(k-k'), \qquad k,k'\in(0,\infty),
\label{toro}
\end{equation}
and are also complete within the subspace of radially symmetric wavefunctions. 

Now $\vert\psi_k\vert^2$, multiplied by the radial volume element  $4\pi r^2{\rm d}r$, yields the radial probability density corresponding to the radially symmetric states $\psi_k(r)$:
\begin{equation}
\rho_k(r)\,{\rm d}r=\frac{2}{\pi}\sin^2(kr)\,{\rm d}r.
\label{seeya}
\end{equation}
Unfortunately $\rho_k(r)$ fails to comply with the cosmological principle as it depends on $r$. However in Eq. (\ref{onda}) below we will find that the actual state of our Universe corresponds to a radial quantum number $k_0=2.9\times 10^{-122}$ natural units. Given the extreme smallness of this value we can approximate $\rho_{k_0}(r)\simeq 2k_0^2r^2/\pi$. Formally this still violates the cosmological principle, but the violation is extremelly small for the following reason. Although $r\in(0,\infty)$, there is a maximum value that $r$ can actually attain, namely the radius of the observable Universe ($R_U\simeq 10^{61}$ natural units). The dimensionless  product $kr$ thus attains a maximum value $k_0R_U\simeq 2.9\times 10^{-61}$, and the quadratic dependence of $\rho_{k_0}(r)$ on $r$ is extremely weak as announced.

\section{Observables in the basis of Laplacian eigenfunctions}\label{beable}

\subsection{The cosmological constant}\label{kosmokonst}

The radial coordinate $r$ in the metric (\ref{metrik}) carries the dimension of length. Hence the operator 
\begin{equation}
\Lambda(r)=\frac{1}{r^2}
\label{landa}
\end{equation}
carries the dimensions of inverse length squared; as such it qualifies as a viable candidate  to represent the cosmological constant; see ref. \cite{MEX} for an identical proposal. 

Next we will obtain the matrix representing the operator (\ref{landa}) within the subspace of spherically symmetric states in Hilbert space. In the basis  (\ref{espan}) we have
\begin{equation}
\langle\psi_{k}\vert\frac{1}{r^2}\vert\psi_{k'}\rangle={\rm min}(k,k'), \qquad k,k'\in(0,\infty).
\label{fucktrump}
\end{equation}
{}For an estimate of the orders of magnitude involved we set $k'=k$. Then
\begin{equation}
\langle\psi_{k}\vert\Lambda\vert\psi_{k}\rangle=k, \qquad k\in(0,\infty).
\label{fuckputin}
\end{equation}
The above expectation value grows linearly with the quantum number $k$, in agreement with the results obtained in ref. \cite{PLANO} in the limit of a Newtonian spacetime. We see that the current value of the cosmological constant is attained for $k=k_0=2.9\times 10^{-122}$. Thus
\begin{equation}
\psi_{k_0}(r)=\frac{1}{\pi\sqrt{2}}\frac{\sin(k_0r)}{r}\simeq\frac{k_0}{\pi\sqrt{2}}, \qquad k_0=2.9\times 10^{-122}\;{\rm natural}\;{\rm units}
\label{onda}
\end{equation}
is the wavefunction describing the matter contents of a flat, FLRW Universe such that the expectation value $\langle\psi_{k_0}\vert\Lambda\vert\psi_{k_0}\rangle$ equals the measured value $\Lambda=2.9\times 10^{-122}$ natural units.\footnote{Despite their numerical coincidence, $k_0$ is an inverse  length while $\Lambda$ is an inverse length squared.}

\subsection{Gravitational Boltzmann entropy}\label{gravbolent}

The conceptual as well as the experimental aspects of the entropy of the Universe have long been the subject of keen attention \cite{ASTROPH, FRAMPTON, ODINTSOV, PENROSE}. In our approach we can continue to apply the arguments put forward in refs. \cite{PLANO, HIPERBOLICO, ESFERICO} and define the operator
\begin{equation}
{\cal S}(r)=MH_0r^2
\label{trumptheloser}
\end{equation}
as a measure of the gravitational Boltzmann entropy of the Universe. We are interested in the matrix $\langle\psi_k\vert {\cal S}\vert\psi_{k'}\rangle$ in the basis (\ref{espan}) of spherically symmetric states. We find 
\begin{equation}
\langle\psi_k\vert r^2\vert\psi_{k'}\rangle=\frac{2}{\pi}\int_0^{\infty}{\rm d}r\,r^2\,\sin(kr)\sin(k'r).
\label{trumpkaka}
\end{equation}
The above integral diverges at the upper limit, but we can regularise it by imposing an infrared cutoff given by the radius $R_U$ of the observable Universe and averaging. Then (\ref{trumpkaka}) becomes
\begin{equation}
\langle\psi_k\vert r^2\vert\psi_{k'}\rangle_{\rm reg}=\frac{2}{\pi R_U}\int_0^{R_U}{\rm d}r\,r^2\,\sin(kr)\sin(k'r)
\label{trumpeatshit}
\end{equation}
where the averaging factor $R_U$ in the denominator is necessary for dimensional reasons. Indeed the unregularised integral in Eq. (\ref{trumpkaka}) carries the dimensions of length squared, but its regularised counterpart in Eq. (\ref{trumpeatshit}) carries the dimensions of length cubed. Evaluating the regularised matrix element we find
\begin{equation}
\langle\psi_k\vert r^2\vert\psi_{k'}\rangle_{\rm reg}\simeq\frac{R_U}{\pi}\left\{\frac{\sin[(k-k')R_U]}{k-k'}-\frac{\sin[(k+k')R_U]}{k+k'}\right\},
\label{trumpmerde}
\end{equation}
after keeping just leading terms in $R_U$. Again we set $k'=k$ in order to estimate the orders of magnitude involved and retain just leading terms in $R_U$:
\begin{equation}
\langle\psi_k\vert {\cal S}\vert\psi_{k}\rangle_{\rm reg}\simeq \frac{MH_0R_U^2}{\pi}.
\label{kicktrump}
\end{equation}
Substituting the numerical values of $M$, $H_0$ and $R_U$ we find
\begin{equation}
\langle\psi_k\vert {\cal S}\vert\psi_{k}\rangle_{\rm reg}\leq 10^{123},
\label{trumpfuckoff}
\end{equation}
regardless of the radial quantum number $k$ and in compliance with the holographic principle.

\section{Discussion}\label{disku}

Understanding the dark Universe is one of the major challenges of current physics. As a contribution to this endeavour, here we have proposed a geometrical model in which a free Schroedinger field mimicks the cosmological fluid within a flat FLRW spacetime. Our aim has been a theoretical computation of the cosmological constant $\Lambda$, under the assumption (\ref{espansione}) of exponential expansion of the scale factor of the FLRW metric.

To begin with, following Madelung \cite{MADELUNG}, we have decomposed the Schroedinger wavefunction into amplitude and phase (see Eq. (\ref{fatto})). This decomposition renders the usual Schroedinger equation strictly equivalent to the set of two equations (\ref{konntt}) and (\ref{quuve}) (the continuity equation for the Schroedinger probability fluid, and the quantum Hamilton--Jacobi equation, respectively). However, the advantage of the Madelung decomposition is that it bijectively maps the Schroedinger probability fluid into a perfect Euler fluid (see Eq. (\ref{tabla})). This is useful because the nonrelativistic cosmological fluid qualifies as a perfect Euler fluid. 

The Schroedinger field $\Psi$ is taken to represent the overall (baryonic plus dark) matter contents of the Universe, assuming this matter is Newtonian (nonrelativistic); however the spacetime is assumed relativistic (flat FLRW space). The coupling between the Einstein field equations and the Schroedinger field $\Psi$ is effected as follows: first, the gravitational field equations are expressed as the Friedmann--Lema\^itre equations (\ref{kklarx}). Then the corresponding matter density $\rho$ and pressure $p$ are expressed in terms of the Schroedinger wavefunction $\Psi$ as given in Eqs. (\ref{fatto}) and (\ref{tabla}).

Under the assumption for the scale factor made in (\ref{espansione}), the Laplacian $\square_{\rm FLRW}$ in Eq. (\ref{laplos}) simplifies to the operator $\square_{H_0}$ in Eq. (\ref{laplass}), which is the Laplacian operator we concentrate on for the rest of the paper. We have succeeded in solving the spectral problem for the operator $\square_{H_0}$; Eq. (\ref{fattoz}) exhibits a complete set of Laplacian eigenfunctions and the corresponding eigenvalues.

The Laplacian eigenfunctions $\Psi(t,r,\theta,\phi)$ factorise as $T(t)\psi(r,\theta,\phi)$, the spatial piece $\psi(r,\theta,\phi)$ being given in Eq. (\ref{doce}). The latter coincide with the usual eigenfunctions $\psi_{klm}(r,\theta,\phi)$ for a free, scalar Schroedinger particle \cite{LANDAU2}. We take these $\psi_{klm}(r,\theta,\phi)$ to span the Hilbert space of stationary quantum states for the nonrelativistic matter contents (baryonic plus dark) of the Universe. Thus our description of the cosmological fluid is given by the probability fluid of a nonrelativistic scalar particle of mass $M$, the latter being the overall (baryonic plus dark) mass of the Universe. The volume density $\rho$ of this fluid in the state $\psi_{klm}$ is given by $\vert\psi_{klm}\vert^2$. In order to comply with the cosmological principle, we have restricted our attention to the subspace of radially symmetric states, {\it i.e.}\/, to the subspace of Hilbert space spanned by the eigenfunctions $\psi_{k00}$ (hereafter denoted more simply $\psi_k$).

We have proposed Eq. (\ref{landa}) as an operator to represent the cosmological constant and obtained the corresponding matrix (\ref{fucktrump}) within the subspace of radially symmetric quantum states. Of course, this procedure has yielded not one value for $\Lambda$ but an infinite--dimensional matrix. The eigenvalue (or eventually the expectation value) that comes closest to the experimentally measured value of $\Lambda$ allows one to identify the radial quantum number $k_0$ of the state $\psi_{k_0}$ that best describes the matter within the actual Universe. It turns out that the best fit is obtained for $k_0=2.9\times 10^{-122}$ (in natural units), which lies extremely close to $k=0$. This latter fact guarantees that the actual wavefunction of the Universe complies with the cosmological principle reasonably accurately. 

A variation on the above theme, given by Eq. (\ref{trumptheloser}), has allowed us to compute the gravitational Boltzmann entropy ${\cal S}$ of the Universe when the matter therein is described by the quantum states $\psi_{k}$; again the result is an infinite--dimensional matrix. As a consistency check we have verified that the upper bound set by the holographic principle \cite{THOOFT, SUSSKIND} is respected by our results. By this we mean that the expectation values $\langle\psi_{k}\vert {\cal S}\vert\psi_{k}\rangle$ never exceed the upper bound ${\cal S}=10^{123}$, regardless of the value of the radial quantum number $k\in(0,\infty)$.

At the request of a referee, we would like to comment briefly on the relationship between our approach to the cosmological constant and that of the string--theory landscape \cite{BENA, ELIZALDE0}. Indeed, string theory exhibits a huge number (possibly some $10^{500}$) of physically inequivalent vacua. Each one of these vacuum states gives rise to a hypothetical Universe with potentially different values of the fundamental physical constants. Among the latter one should count the cosmological constant $\Lambda$. This wealth of different values for $\Lambda$ is reminiscent of the eigenvalue spectrum $(0,\infty)$ for our operator $\Lambda$ in  Eq. (\ref{landa}). In our approach, this continuum of possible eigenvalues for the cosmological constant is unconstrained; we are satisfied to find that the experimentally measured value of $\Lambda$ is actually realised as one of the allowed eigenvalues of an operator. This situation is mimicked in the string--theory landscape, where the sheer size of $10^{500}$ clearly allows one to regard the different possible values of $\Lambda$ as forming a continuum. No anthropic principle is invoked in our approach; in the absence of a guiding principle to make a prediction, we simply pick the eigenvalue that best fits the experiments. All this notwithstanding, it is remarkable that the simple quantum--mechanical model presented here shares some features in common with the more sophisticated constructions of string theory.

In upcoming work we propose to analyse the cosmological constant within the framework of causal fermion systems \cite{FINSTER1}, along the lines of the already existing mechanism for baryogenesis \cite{FINSTER2}. Our approach shares the geometrical point of view of refs. \cite{COGNOLA, MATONE1, MATONE2}. Our results are also in line with thermodynamical approaches such as those of refs. \cite{JACOBSON, PADDY2, PADDY3, VERLINDE1, VERLINDE2}, where spacetime is argued to be a derived concept and the Einstein field equations arise as a thermodynamical equation of state. We hope to report on these issues in the future.

\end{document}